\documentclass[12pt]{article}
\usepackage{amssymb}
\begin{document}


\title{Note on a Pattern from CP Violation in Neutrino Oscillations}
\author{
R.~G.~Moorhouse$^b$\\[.4cm]
\small $^b$University of Glasgow,Glasgow G12 8QQ, U.K.\\
}
\maketitle
\begin{abstract}
\noindent

 A virtual graphical construction is made to show the difference between 
 neutrino and anti-neutrino oscillations in the presence of CP
 violation with CPT conservation. 

\end{abstract}

\section{Introduction}

 There is interest in the possibility that CPT violation may occur 
 and then show in neutrino oscillation experiments\cite{BL},\cite{CDK},
 \cite{MINOS}.. However this may be,
 CP violation is long established and  it is of importance
 to seek it in neutrino oscillation results. If one takes a conservative
 point of view that nature conserves CPT, and there are 3 generations
 of neutrinos, then the consequences of CP 
 violation in neutrino oscillation become more definite. In particular
 if CP were conserved $\nu$ transition  probabilities would be the same 
 as $\bar{\nu}$ transition  probabilities while the occurence therein of
 CP violation makes these different \cite{Kayser}. This differencc is
 dependent on the neutrino oscillation parameter $L/E$, $L$
 being the distance of travel from creation to detection and $E$ the
 energy of the initial neutrino: and it is also sensitive to the value of
 the small ratio of neutrino mass squared differences. There results a
 complicated 2-variable dependence in addition to the linear 
 dependence on the leptonic Jarlskog parameter, $J_{lep}$\cite{HSW}. 

\section{Pattern for the difference between $\nu$ and $\bar\nu$
 oscillations}
 The input to the formula for neutrino transition probabilities is
 largely from the mixing matrix elements $U_{\alpha i}$ where $\alpha$
 is one of the 3 flavour indices and $i$ one of the 3 mass eigenstate
 indices. From these 9 elements plaquettes \cite{BD} can be constructed,
 these being phase invariant products of 2 $U$ elements multiplied
 by products of 2 $U^{\star}$ elements which occur in transition
 probabilities $\nu_{\alpha} \to \nu_{\beta}$ of neutrino beams.
 Here $\alpha,\beta$ are flavour indices of beam neutrinos. The
 construction is as follows.

 Greek letters denoting flavour indices $(e,\mu,\tau)$ and Roman letters
 mass eigenstate indices there are 9 plaquettes,labelled
 $\Pi_{\alpha i}$:  
\begin{equation}
 \Pi_{\alpha i} \equiv 
 U_{\beta j}U^{\star}_{\beta k}U_{\gamma k}U^{\star}_{\gamma j}
\label{119}
\end{equation}
 where $\alpha,\beta,\gamma$ are non-equal and in cyclic order and
 $i,j,k$ are also non-equal and in cyclic order (The pattern discussed in 
 this paper applies for an inverted hierarchy as well as for the 
 normal hierarchy; that is there is no necessary association 
 between a particular $\alpha$ and a particular $i$.) .

 Making use of well-known formalism \cite{Kayser} the beam transition
 probability for $\nu_{\alpha} \to \nu_{\beta}$,$\alpha \neq \beta$
 can be written as
 \begin{eqnarray}
 P(\nu_{\alpha} \to \nu_{\beta})= 
 -4\sum_{i=1}^3 \Re(\Pi_{{\gamma}i})\sin^2((m_{{\nu}k}^2-m_{{\nu}j}^2)
  L/4E)\\
 +2\sum_{i=1}^3 \Im(\Pi_{{\gamma}i})\sin((m_{{\nu}k}^2-m_{{\nu}j}^2)
  L/2E)\label{120}
\end{eqnarray}

where L is the length travelled by neutrino energy E from creation
to annihilation at detection. The survival probability,
$P(\nu_{\alpha} \to \nu_{\alpha})$, (given in \cite{Kayser}) can be
 calculated from the transition probabilities above. So the
 $3 \times 3$ plaquette matrix  $\Pi$ and the neutrino mass
 eigenstate values squared differences carry all the information on 
transition and survival probabilities of a given beam.

 The last term (\ref{120}) being only non-zero when CP is not conserved.
 Indeed all the nine $\Im \Pi_{{\alpha} i}$ are equal and equal to \cite{HDS}
 the leptonic Jarlskog invariant; $$\Im \Pi_{{\alpha} i}=J_{lep}$$.
 So J, as usual,
 is signalling CP violation. With CPT invariance the transition probability 
 for anti-neutrinos $P(\bar{\nu}_{\alpha} \to \bar{\nu}_{\beta};\Pi )=
 P(\nu_{\alpha} \to \nu_{\beta};\Pi^\star)$  \cite{Kayser}. Thus the 
 contribution of CP violation in anti-neutrino transitions is of the same 
 magnitude but opposite sign to that in neutrino transitions, giving rise to 
 a, in principle measurable, difference in the overall probability since the
 CP conserving contributions are the same..

 The part of the probability (\ref{120}) arising from CP violation is 
 2$J\xi$ where
 \begin{equation}
 \xi= \sum_{i=1}^3 \sin((m_{{\nu}k}^2-m_{{\nu}j}^2)L/2E)
\label{121}
\end{equation}
 This sum of sine functions (the sum of whose arguments is zero)
 may readily be transformed to
 \begin{equation}
 \xi= 4\sin(x_d)\sin(y_d)\sin(x_d+y_d) 
\label{122}
\end{equation}
 \begin{eqnarray}
 (x_d,y_d)=(d_1L/4E,d_2L/4E)  \\
 d_1=(m_{{\nu}2}^2-m_{{\nu}1}^2), 
 d_2=(m_{{\nu}3}^2-m_{{\nu}2}^2)\label{123}
\end{eqnarray} 
.
  Now consider the function
 \begin{equation}
 \Xi(x,y) \equiv 4\sin(x)\sin(y)\sin(x+y) 
\label{126}
\end{equation}
 where in $\Xi (x,y)$ the arguments $x,y$ are freely varying 
 and not restricted as in $\xi$. This function $\Xi$ has multiple
 maxima,minima with values $+3\sqrt3/2,-3\sqrt3 /2$
 at arguments say $x_m$ and $y_m$ which are integer multiples of
 $\pi/3$ (but obviously not spanning all such integer multiples).

 Given mass squared differences then $\xi(L/E)$ is a function varying
 only with $L/E$ and the above maxima and minima cannot generally be
 attained. However one can distinguish regions of $L/E$ where relatively 
 high values of $\xi$ are attained. These are, naturally, given by values
 of $x_d,y_d$ near to $x_m,y_m$ points of $\Xi$. These latter points
 can be located in the $(x,y)$ plane through the necessary condition
 that there the first derivatives of $\Xi$ with respect to both $x$ and $y$
 should vanish. A simple geometrical picture can be given as follows.

 On the $x$ and $y$ positive quartile of the plane construct a square grid
 with neighbouring grid lines a distance $\pi/3$ apart resulting in a pattern 
 of squares of side $\pi/3$. All the maximum and minimum points of $\Xi$
 are at intersection points of the grid lines and are given by
 \begin{eqnarray}
 (x_m,y_m)=(1+3l,1+3k)\pi/3 \label{127}\\
 (x_m,y_m)=(2+3l,2+3k)\pi/3 \label{128}
\end{eqnarray}   
 where $l$ and $k$ are any non-negative integers. The points $(\ref{127})$
 have $\Xi=3\sqrt3/2$ and the points $(\ref{128})$ have $\Xi=-3\sqrt3/2$.

 It is near these special points in the $x,y$ plane that $\xi(L/E)$ (eqn.
 \ref{121}) has numerically large values. Note that for seeking observation
 of CP violation using the difference between $\nu$ and $\bar{\nu}$
 transitions it does not matter whether $\xi$ is positive or negative
 so both maximum and minimum points of $\Xi$ are equally potentially
 important. 

 As $L/E$ varies the points $(x_d,y_d)$ (\ref{123}) trace a straight
 line in the $(x,y)$ plane starting at $(0,0)$  and ascending as $(L/E)$
 increases. This line of $\xi$ makes a small angle arctan$(d_1/d_2)$ with the
 y-axis and pases through the archipelago of special points given by
 (\ref{127},\ref{128}). Points $(x_d,y_d)$ on the line of $\xi$ which are 
 close to the $\Xi$  special points (\ref{127},\ref{128}) give numerically
 large values of $\xi$ and the associated values of $L/E$ signify neutrinos
 whose transtions contain a relatively large CP violating part. To give an
 idea of how much of the plane has a value of $\Xi$ 
 near maximum or minimum then the value of $\Xi$ near the special points 
 should be evaluated.  Let $\delta$ be the distance between a near point
 and the special point which it is near to. Then near a maximum or minimum
 $\Xi=\pm 3\sqrt3/2(1- \Delta)$ respectively. where $\Delta \leq 2\delta^2$ .
. Thus within an area limited by  $\delta = .2$ (noting that a grid 
 square has sides length $\pi/3$) the value of $\Xi$ is nearly equal to that 
 at the special grid point.

 So the structure of $\Xi$ is such that for certain intervals (not large)
 of $L/E$ the contribution of $J\xi$ to CP violation in neutrino transitions
 (measured by the difference between $\nu$ and $\bar{\nu}$ transitions) is
 much bigger than an average over larger intervals.
 Such an interval of $L/E$ is when the line of
  $L/E$ passes near a peak or trough of $\Xi$. The example that follows is
  only illustrative, though by happenstance rather striking. There are,
 obviously, uncertaities in the prescription due to considerable relative
 uncertainties in the value
 (though certainly small) of $d_1/d_2$ and also in the different plaquette
 values found in different theories.
 
    Take  $$d_1=8.0\times 10^{-5} eV^2, d_2=2.5\times 10^{-3} eV^2, $$
 so that $d_1/d_2=.032, d_2/d_1=31.25$. Consider the grid line $y=62\pi/3$
 (given by $k=20$ eqn.\ref{128}). On this grid line $\Xi$ has a minimum value
 $-3\sqrt3/2$ at $x=2\pi/3$ and the line of $\xi$ crosses the grid line at
 $x=(2\pi/3- .016)$ which means that $\xi$ has almost attained the minimum
 value, $-3\sqrt3/2$,and that $L/4E = 62\pi/3d_2 = 0.248 \times 10^5\pi/3$.

 The values of the plaquettes defined from the MNS mixing matrix
 elements, $U_{\alpha i}$,depend on the phases of these elements. A
 theory can give the moduli and phases of these elements and the 
 consequent plaquette have the virtue of being obviously invariant
 under phase redefinitions of the eigenstates $\nu_{\alpha}$ and of the
 eigenstates $\nu_i$. In the absence of experimental data on the phases of
 the $U_{\alpha i}$, but the existence of some considerable experimental 
 guidance on the moduli it seems not unreasonable to use, as a specimen
 for present purposes, one of the theories which gives a matrix of
 moduli squared
 resembling that of the tri-bimaximal mixing hypothesis \cite{HPS}.
 The particular theory used \cite{M}
 incorporates the values of the mass squared differences 
 given and used above and has the matrix of MNS modulus squared elements:.
\begin{equation}
 \left[\begin{array}{ccc}
.638 & .344 & .017\\
.260 & .331 & .409\\
.102 & .325 & .573
\end{array}\right], 
 \label{132}
 \end{equation}
 bearing a distinct resemblance to the postulated 'ideal' structure 
 of this matrix in tri-bimaximal mixing \cite{HPS}: (The theoretical
 model \cite{M} produces the MNS matrix using the normal hierarchy.)

 As previously noted $\Im \Pi_{{\alpha} i}=J_{lep}$
 (for all 9 elements of $\Pi$) and the contribution of CP violation to the 
 transition probabilities is given by 
 \begin{equation}
 P_{CP}(\nu_{\alpha} \to \nu_{\beta})=2\xi(L/E)J_{lep}  
\label{129}
\end{equation}  
 for all $\alpha$ not equal to $\beta$.

 For this particular model $J_{lep}=.01744$ and at the nearly minimum
 point discussed above
 \begin{eqnarray}
P(\bar{\nu}_\mu \to \bar{\nu}_e)=P(\nu_e \to \nu_\mu) = .3470\label{130}\\
P(\nu_\mu \to \nu_e)=P(\bar{\nu}_e \to \bar{\nu}_\mu) = .5276\label{131}
\end{eqnarray}   
 the difference of 0.1806. It should be emphasized that this large
 difference depends not only on the nearness to a special point, 
which is a concept independent of the any particular model of the MNS
 matrix but also on the particular model having a maybe atypically 
 large value of $J_{lep}$. Naturally this type of experimental 
 comparison may yield some knowledge of $J_{lep}$.

 The numerical value of $L/E$ given above in units $ev^{-2}$ may, on 
 inserting the appropriate dimensionful value of $\hbar c$, 
 be related to the experimental conditions through 
 $$L/4Eev^{-2}=1.266L(km)/E(Gev)=0.248 \times 10^5\pi/3$$.
 Values of $L/E$ of this order may be appropriate for atmospheric muon
 neutrinos created on the opposite side of the earth to the detector. 
 However the large value of $L/E$ highlights the probable accelerator
 experiment  difficulties
 of getting near to some of the special grid points of $\Xi$; this
 sharply depends on the precise value of the gradient of the line of $\xi$.

 It is clear that there is at present no reliable prediction of detailed
 exprimental results; most importantly because the precise slope of the
 line of $\xi$ in the $(x,y)$ plane is uncertain, being the ratio of
  neutrino mass differences. Rather, any value of the construction
 is that experiment may give information on its physically significant
 parameters. Information may of course be obtained by computer
 evaluation of the transition probabilities (\ref{120}) for 
 very many multiple parameter choices, diligent attention enabling 
 the construction of cognitive or computer maps.

 The author thanks David Sutherland and Colin Froggatt for comments
 on this note..


\newpage
\begin{thebibliography}{8}
\bibitem{BL}G.Barenboim and J.D.Lykken, arXiv: 0908.2993[hep-ph]
 G.Barenboim, L Borissov and J.D.Lykken, arXiv: 0212.116[hep-ph]
 \bibitem{CDK}D.Choudhury,A. Datta and A.Kundu arXiv:1007.2923[hep-ph].
 \bibitem{MINOS} P. Adamson et al.arXiv:1007.2791[hep-ph].
\bibitem{Kayser} B.Kayser 'Neutrino mass, mixing and flavor' in
 2008 Review of Particle Physics: C. Amaler et al.,Physics Letters B667,
 1(2008).
\bibitem{HSW}P. F. Harrison, W. G. Scott and T. J. Weiler,
 arXiv: 0908.2993[hep-ph]
\bibitem{BD} J. D. Bjorken and I. Dunietz,Phys. Rev. D36, 2109 (1987).
 \bibitem{HDS} P F Harrison, S Dallison and W G Scott, arXiv:0904.3071[hep-ph]
\bibitem{HPS}P. F. Harrison, D. H. Perkins and W. G. Scott,
 Phys.Lett. B530, 167 (2002).
 \bibitem{M} R. G. Moorhouse, Phys. Rev. D77,053008 (2008)

\end{thebibliography}
\end{document}